\documentclass[prl,twocolumn,superscriptaddress, unsortedaddress, amsfonts,amssymb,amsmath,titlepage]{revtex4} 

\usepackage[utf8]{inputenc}
\usepackage[hidelinks, breaklinks=true]{hyperref}
\usepackage{latexsym}
\usepackage{verbatim} 
\usepackage{natbib}
\usepackage{graphicx}
\usepackage{textcomp}
\usepackage{xcolor}
\usepackage{xspace} 
\usepackage{ulem} 
\usepackage{siunitx}
\usepackage{currfile}
\usepackage{fdsymbol}
\usepackage{mathrsfs}
\usepackage{CJKutf8} 

\newcommand{\pgi}{Peter Grünberg Institut (PGI-3), Forschungszentrum Jülich, 52425 Jülich, Germany}
\newcommand{\jara}{Jülich Aachen Research Alliance (JARA), Fundamentals of Future Information Technology, 52425 Jülich, Germany}
\newcommand{\diam}{Diamond Light Source Ltd, Didcot, OX110DE, Oxfordshire, United Kingdom}
\newcommand{\rwth}{Experimentalphysik IV A, RWTH Aachen University, Otto-Blumenthal-Straße, 52074 Aachen, Germany}

\newcommand{\myref}[7]{\href{http://dx.doi.org/#7}{#1, #2, #3 \textbf{#4}, #5 (#6).}} 
\newcommand{\tB}{$30^\circ$-tBLG\xspace}
\newcommand{\Gnul}{G-$R0^\circ$\xspace}

\newcommand{\Gdrei}{G-$R30^\circ$\xspace}
\newcommand{\hBNnul}{hBN-$R0^\circ$\xspace}
\newcommand{\ZL}{ZLG-$R30^\circ$\xspace}
\newcommand{\sixSqrt}{$(6\sqrt{3}\times6\sqrt{3})R30^\circ$\xspace}
\newcommand{\dxd}{$(3\times 3)$\xspace}
\newcommand{\C}{$^\circ$C\xspace}
\newcommand{\BZ}{$\%\mathrm{BZ_{SiC}}$\xspace}
\newcommand{\Gst}{G-1st\xspace}
\newcommand{\Gnd}{G-2nd\xspace}

\begin{document}
\title{Surfactant-Mediated Epitaxial Growth of Single-Layer Graphene in an Unconventional Orientation on SiC}

\author{F.C.~Bocquet}   \email{f.bocquet@fz-juelich.de} \affiliation{\pgi} \affiliation{\jara}
\author{Y.-R.~Lin (\begin{CJK*}{UTF8}{bsmi}林又容\end{CJK*})}      \affiliation{\pgi} \affiliation{\jara} \affiliation{\rwth}
\author{M.~Franke}      \affiliation{\pgi} \affiliation{\jara} \affiliation{\rwth}
\author{N.~Samiseresht} \affiliation{\pgi} \affiliation{\jara} \affiliation{\rwth}
\author{S.~Parhizkar}   \affiliation{\pgi} \affiliation{\jara}
\author{S.~Soubatch}    \affiliation{\pgi} \affiliation{\jara}
\author{T.-L.~Lee (\begin{CJK*}{UTF8}{bsmi}李天麟\end{CJK*})}      \affiliation{\diam}
\author{C.~Kumpf}       \affiliation{\pgi} \affiliation{\jara} \affiliation{\rwth}
\author{F.S.~Tautz}     \affiliation{\pgi} \affiliation{\jara} \affiliation{\rwth}

\begin{abstract}
  We report the use of a surfactant molecule during the epitaxy of graphene on SiC(0001) that leads to the growth in an unconventional orientation, namely $R0^\circ$ rotation with respect to the SiC lattice. It yields a very high-quality single-layer graphene with a uniform orientation with respect to the substrate, on the wafer scale. We find an increased quality and homogeneity compared to the approach based on the use of a pre-oriented template to induce the unconventional orientation. Using spot profile analysis low energy electron diffraction, angle-resolved photoelectron spectroscopy, and the normal incidence x-ray standing wave technique, we assess the crystalline quality and coverage of the graphene layer. Combined with the presence of a covalently-bound graphene layer in the conventional orientation underneath, our surfactant-mediated growth offers an ideal platform to prepare epitaxial twisted bilayer graphene via intercalation.
\end{abstract}

\maketitle
Since the isolation of graphene (G) as the first two-dimensional material by micromechanical cleavage of graphite in 2004 \cite{Novoselov2004}, efforts have been made to control its electronic properties without affecting the lattice. In 2007, twisted bilayer graphene (tBLG) has already been identified as a case in point \cite{Lopes2007}. In this material, the twist angle \cite{Rozhkov2016} as well as the relative strain \cite{Huder2018, Naumis2017, Kumar2015, Beechem2014} between the top and the bottom layers are decisive parameters that determine the electronic properties. Among the intriguing properties of tBLG are chirality \cite{Stauber2018, Morell2017, Kim2016}, magnetism \cite{Sboychakov2018, Gonzalez2017}, and a tunable band gap \cite{Rozhkov2017, Liu2015, Muniz2012}. More recently, tBLG has attracted additional interest after unconventional superconductivity was discovered for twist angles of approximately 1.1$^\circ$ \cite{Cao2018_article, Cao2018_letter, Yankowitz2019, Yoo2019}. For large twist angles, near 30$^\circ$, electrons are expected to localize, giving rise to flat bands and strong correlation physics \cite{Pal2019, Pal2014, Moon2019, Park2019}. Also, large twist angle tBLG is predicted to show nontrivial higher-order band topology \cite{Park2019b}.

Up to now, most experimental studies of tBLG are performed on manually stacked, exfoliated graphene single layers. This method has the advantage that virtually any material can be stacked with any angle, but the moderate material quality, small sample size and limited twist angle accuracy are its major shortcomings. For example, the presence of interlayer contaminants forming bubbles is a challenging issue \cite{Frisenda2018}. Moreover, this method is not scalable. In contrast, epitaxial growth is more limited in scope, but if the correct growth parameters for a given stack and a given twist angle have been identified, it is reproducible, accurate, and readily scalable, and it offers unrivaled cleanness and control at the atomic scale. 

tBLG can be grown epitaxially on metals. Examples are Pt(111) \cite{Yao2018}, Pd foil \cite{Zuo2018}, Pd(111) \cite{Murata2012}, Cu foil \cite{Peng2017, Hu2017, Lu2013},  Ni--Cu gradient foil \cite{Gao2018}, Ir(111) \cite{Nie2011}, and Ni(111) \cite{Iwasaki2014}. Thermal decomposition of 6$H$-SiC(000$\bar{1}$) (C-face) is another route to obtain tBLG  \cite{Lee2011, Tejeda2012, Razado2016}. However, in all of these cases one finds random twist angles and/or random orientations across the sample, and each domain of a given twist angle has a typical maximum diameter of 1 to 10~$\mu$m.

In contrast, high-quality wafer-size quasi-freestanding bilayer  graphene \textsl{without twist} can be grown epitaxially on the wide band-gap semiconductor SiC(0001) (Si-face) by thermal decomposition, to produce the epitaxial monolayer graphene (EMLG), and subsequent atomic intercalation \cite{Riedl2009, Emtsev2009}. In EMLG, the graphene layer is oriented 30$^\circ$ with respect to the SiC lattice (\Gdrei) and lies on the so-called zeroth-layer (\ZL, with  \sixSqrt superstructure) that consists of a graphene layer with the same orientation but covalently bound to SiC. Via intercalation, e.g., with hydrogen atoms, the SiC is passivated, the \ZL transforms into \Gdrei forming a graphene bilayer without twist. By choosing the intercalated atomic species (even volatile elements), one can tune the interaction strength to the substrate, adjust the doping level (when heavily $n$-doped, the Fermi level can reach the Van Hove singularity of the $\pi$-band), and even produce ballistic bipolar junctions \cite{Riedl2009, Gierz2010, McChesney2010, Virojanadara2010, Walter2011, Emtsev2011, Yagyu2014, Sforzini2015, Baringhaus2015, Rosenzweig2019, Briggs2019, Wolff2019}.

In this Letter, we report a method for epitaxially growing a differently oriented single-layer graphene on the zeroth-layer (namely \Gnul/\ZL) in contrast to the conventional EMLG (\Gdrei/\ZL). It shows high crystalline quality, high coverage, and has a well-defined orientation on a macroscopic scale. Specifically, we anneal the 6$H$-SiC(0001) surface in a borazine (B$_3$N$_3$H$_6$) atmosphere. Thereby, borazine acts as a surface-active molecule (surfactant), enforcing an unconventional orientation of the graphene layer but leaving the conventional \ZL orientation intact. Via intercalation of \Gnul/\ZL, high quality quasi-free-standing epitaxial \tB is obtained. Here, we focus on \Gnul/\ZL, as it is the key to realize \tB, the intercalation being a well-established procedure.

Our method is based on the fact that single-layer hexagonal boron nitride hBN grows epitaxially with $R0^\circ$ orientation (\hBNnul) on 6$H$-SiC(0001) in a borazine atmosphere \cite{Shin2015}. By annealing \hBNnul to higher temperatures in ultra-high vacuum (UHV), a graphene layer gradually replaces \hBNnul, preserving the $R0^\circ$ orientation.  This observation, in conjunction with the known fact that multilayer \Gdrei can be grown on SiC by thermal decomposition, suggests that \tB may be grown on SiC(0001) with the help of hBN \cite{Ahn2018}. However, it is clear that in such a scheme, the quality of the final \tB is limited by the quality of the hBN layer. Structural details about the as-grown and UHV-annealed \hBNnul are given in the Supplemental Material \cite{SuppMat} together with details of the growth process and of the employed characterization methods. In order to eliminate this influence, we use borazine as a surfactant molecule during the growth of \Gnul, thus avoiding the stabilization of a static hBN layer but keeping the influence of hBN nuclei in acting as a template for the growth of graphene in an unconventional orientation. This dynamic approach may offer the advantage that the detrimental effects of domain boundaries in large-scale static hBN, lattice mismatch between graphene and hBN, etc., are avoided. As we show in this Letter, the surfactant-mediated growth indeed yields a single graphene layer in an unconventional orientation with high quality and coverage on top of the conventional \ZL.

\begin{figure}[t!]
  \centering
  \includegraphics[width=.9\columnwidth]{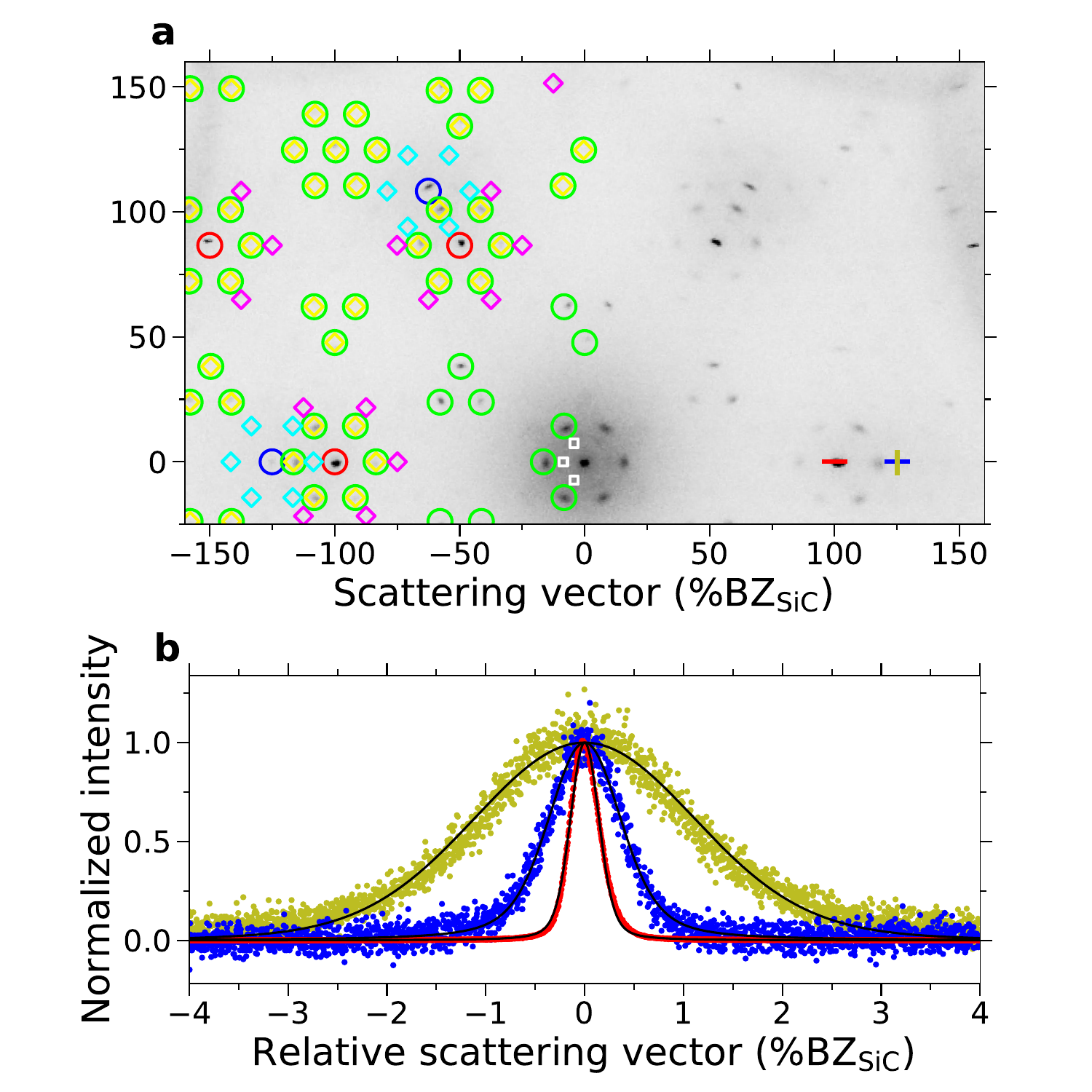}
  \caption{(a)~Distortion-corrected SPA-LEED pattern of S-graphene. Diffraction spot positions were simulated for each of the three structures ($\bigcirc$; SiC(0001) -- red, \ZL -- green, \Gnul -- blue) and for multiple electron diffraction by two  ($\Diamond$) and three ($\square$) structures (see color code in \cite{SuppMat}). (b)~Typical radial profile of the \{10\} SiC spot (red) as well as radial (blue) and azimuthal (olive green) profiles of the \{10\} spot of \Gnul. Colored dots represent experimental data, black lines represent the best fits. The paths used for the spot profiles are indicated in (a) as colored lines.}
  \label{fig:Profile}
\end{figure}

Fig.~\ref{fig:Profile}(a) displays the spot profile analysis low energy electron diffraction (SPA-LEED) pattern of the \textsl{surfactant-induced unconventionally oriented graphene} (S-graphene), which is obtained by annealing the Si-rich SiC(0001)-\dxd reconstruction directly to 1330\C in a borazine atmosphere ($1.5\times 10^{-6}$~mbar). 
The pattern is similar to the one obtained by annealing a \hBNnul template layer to 1300\C in UHV (\textsl{template-induced unconventionally oriented graphene} or T-graphene), see Fig.~S1(c) and Fig.~S2 in \cite{SuppMat}. However, for S-graphene the spots are narrower and we observe more \ZL higher-order diffraction (from which only the (6$\times$6) spots are visible -- green circles $\bigcirc$) and multiple electron scattering ($\Diamond$, $\square$) spots. Specifically, we stress that spots marked with the magenta and cyan diamonds stem from double electron scattering involving the \Gnul reciprocal vectors (see Table~S1 in \cite{SuppMat}) and cannot be accounted for by scattering on a single structure. Fitting the corrected diffraction pattern yields lattice parameters of 32.1$\pm$0.4~Å  for \ZL and 2.45$\pm$0.04~Å  for \Gnul. Importantly, the diffraction pattern of S-graphene can be recovered by a mild annealing in UHV after two months of air exposure. It indicates that this preparation method yields a single layer \Gnul on \ZL with high crystalline quality and coverage, which protects the SiC surface against oxidation.

Deeper insight into the quality of the crystalline structures is provided by the analysis of the respective \{10\} diffraction spot profiles (Fig.~\ref{fig:Profile}(b)). The full widths at half maximum  $w$ of the radial spot profiles arise from the combined effects of the finite instrumental resolution and the finite size of crystalline domains. Hence, $2\pi/w$ represents a lower limit to the average domain size. An azimuthal profile broader than the radial profile is a direct indication of azimuthal disorder. To estimate the azimuthal distribution, assumed in this particular case to be Gaussian, we convolve the radial profile fit result (Voigt) with a Gaussian to fit the azimuthal profile. The fitted standard deviation $\sigma$ of the Gaussian is a measure of the azimuthal disorder. The \{10\} SiC spots (red) have a circular shape and the radial profile is fitted by a Voigt function with $w=(0.341\pm 0.004)$~\BZ. In contrast, the \{10\} radial spot profiles of \Gnul (blue) are broader, $(0.863\pm0.009)$~\BZ, indicating a smaller average domain size than the average SiC terrace size. Moreover, the \{10\} spots of \Gnul are elongated in the azimuthal direction, corresponding to an azimuthal disorder with the standard deviation $\sigma=(0.458\pm 0.002)^\circ$.

\begin{figure}[t!] 
  \centering
  \includegraphics[width=\columnwidth]{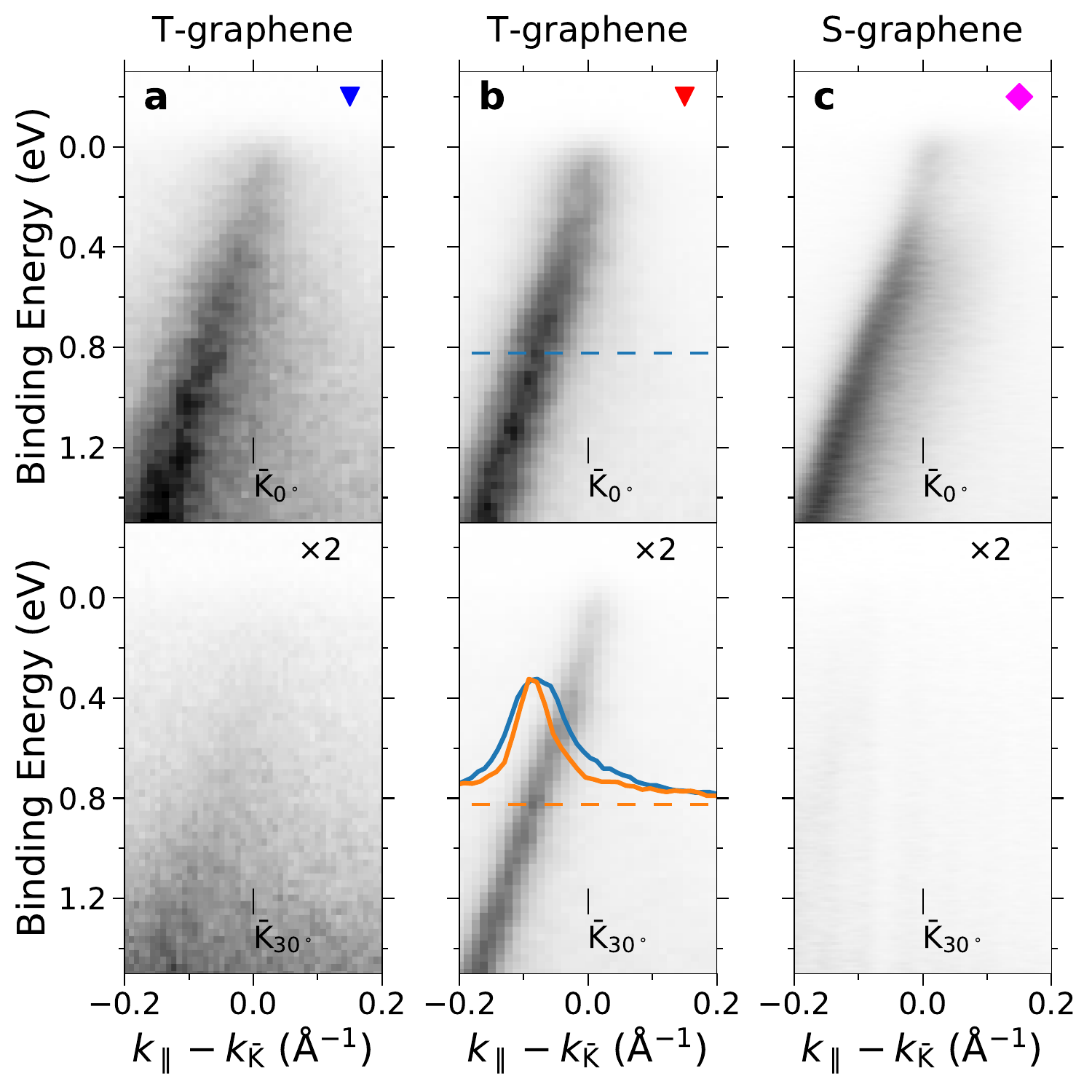} 
  \caption{ARPES EDMs taken around $\mathrm{\bar K}_{0^\circ}$ (upper panel) and $\mathrm{\bar K}_{30^\circ}$ (lower panel, color scale magnified by a factor of two) with $h\nu=110$~eV for T-graphene prepared at (a)~1270\C and (b)~at 1330\C (75$^\circ$ incidence angle), and (c) for S-graphene prepared at 1330\C (87$^\circ$ incidence angle). Normalized momentum distribution curves taken at  a binding energy of about 0.8~eV (dashed lines) around $\mathrm{\bar K}_{0^\circ}$ (blue) and $\mathrm{\bar K}_{30^\circ}$ (orange) are shown as solid lines in (b).}
  \label{fig:ARPES}
\end{figure}

To track the presence of graphene in either azimuthal orientation and to assess its quality and homogeneity, we perform angle-resolved photoemission spectroscopy (ARPES) on samples prepared at different temperatures. Energy distribution maps (EDMs) taken around $\mathrm{\bar K}_{0^\circ}$ (stemming from \Gnul) and $\mathrm{\bar K}_{30^\circ}$ (from \Gdrei) are shown in Fig.~\ref{fig:ARPES}(a) for T-graphene prepared at 1270\C, and in Fig.~\ref{fig:ARPES}(b) for the T-graphene prepared at 1330\C. The intensity gray scale around $\mathrm{\bar K}_{30^\circ}$ is enhanced by a factor of two with respect to the data around $\mathrm{\bar K}_{0^\circ}$. With increasing preparation temperature, the signal-to-background ratio as well as the intensity of the Dirac cone of \Gdrei relative to \Gnul increases and the full width at half maximum (FWHM) of the $\pi$-band decreases.

In contrast, S-graphene shows a Dirac cone around $\mathrm{\bar K}_{0^\circ}$ of comparable quality as for the high preparation temperature side of T-graphene, but no intensity is found around $\mathrm{\bar K}_{30^\circ}$ (see Fig.~\ref{fig:ARPES}(c) and also Fig.~S5(c) in \cite{SuppMat}). Moreover, the EDMs remain unchanged for preparation temperatures ranging from 1300 to 1360\C, thus only the one recorded at 1330\C is shown. For T-graphene, a reduced coverage of \Gdrei comes along with a broad $\pi$-band for \Gnul, because of the lower preparation temperature (Fig.~\ref{fig:ARPES}(a)). For S-graphene, a sharp $\pi$-band can be achieved for \Gnul, without growing \Gdrei, proving its improved homogeneity and quality.

Interestingly, for T-graphene, the FWHM of the $\pi$-band is smaller around $\mathrm{\bar K}_{30^\circ}$ than around $\mathrm{\bar K}_{0^\circ}$ (see colored curves in Fig.~\ref{fig:ARPES}(b)). We note, however, that for both \Gnul and \Gdrei in either T- or S-graphene, the FWHM of the $\pi$-band remains larger than for EMLG \cite{Sforzini2016} and that the Dirac point is located at about 200~meV below the Fermi level (see also \cite{SuppMat}). Two facts can explain these observations. First, doping by substitutional N and B atoms is observed in x-ray photoemission spectroscopy (XPS), and it is known to lead to a band broadening \cite{Sforzini2016}. Second, the sizeable azimuthal distribution observed in SPA-LEED for the \Gnul layer in S-graphene (Fig.~\ref{fig:Profile}) and T-graphene (Fig.~S1(c) in \cite{SuppMat}) samples can also contribute to the broadening around $\mathrm{\bar K}_{0^\circ}$.

\begin{figure}[t!]
  \centering
  \includegraphics[width=\columnwidth]{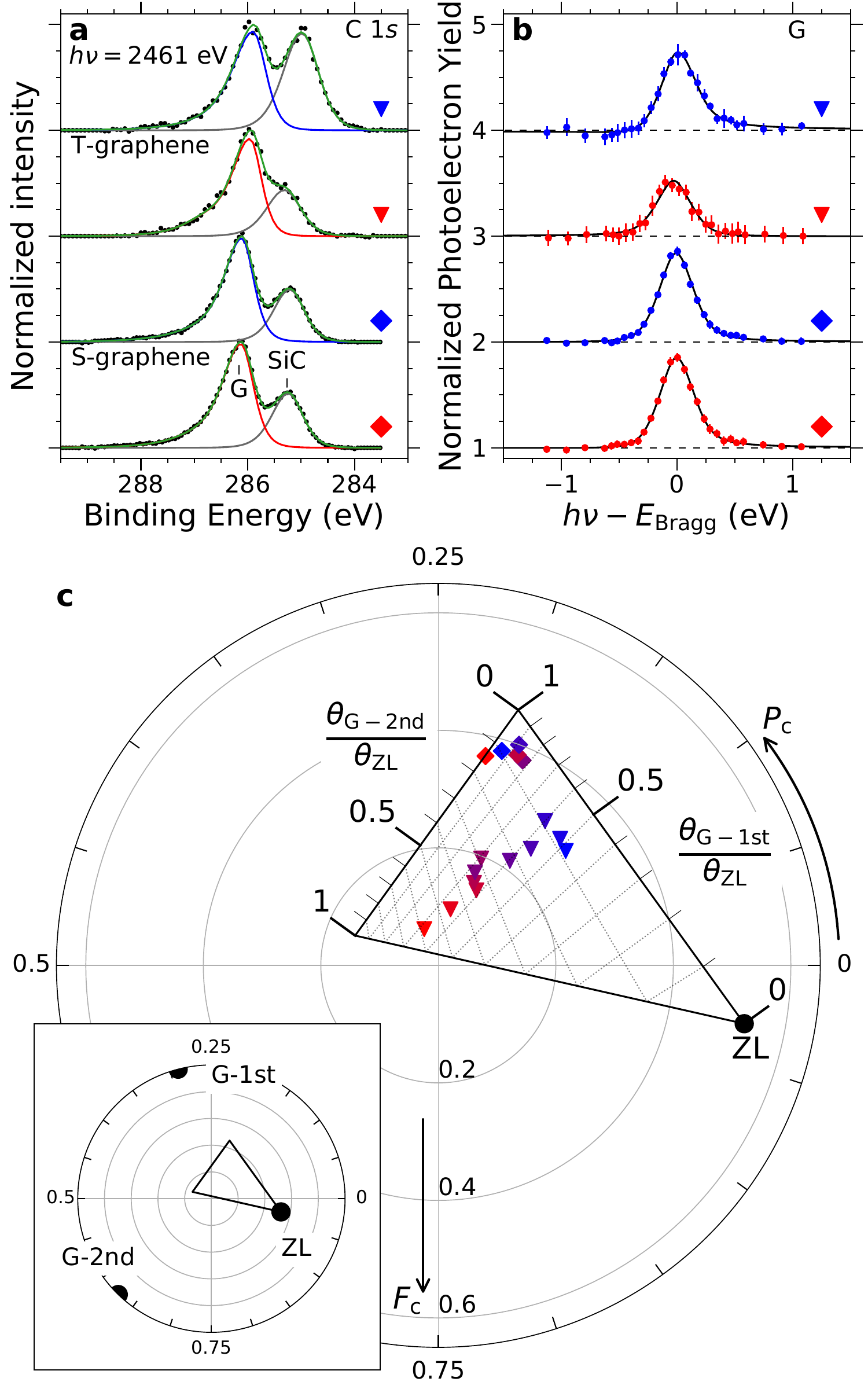}
  \caption{(a)~C~$1s$ core-level spectra taken off-Bragg with a two-component fit model for T-graphene and S-graphene. For T-graphene, the color code blue to red corresponds to a preparation temperature ranging from 1270 to 1330\C, while for S-graphene it corresponds to the range 1300 to 1360\C. (b)~The corresponding yield curves of the high-binding-energy component (G), shifted for clarity. The colored points are the data; the black lines are the fits. (c) Argand diagram summarizing all NIXSW fit results obtained from T-graphene ($\medblacktriangledown$) and from S-graphene ($\medblackdiamond$)  (the preparation temperature is represented by the symbol color). The large black full circles in the inset represent the expected vectors $\mathcal{D}_\mathrm{ZL}$, $\mathcal{D}_\mathrm{G-1st}$ and $\mathcal{D}_\mathrm{G-2nd}$ and constitute the anchoring points of the additional axes $\theta_\mathrm{G-1st}/\theta_\mathrm{ZL}$ and $\theta_\mathrm{G-2nd}/\theta_\mathrm{ZL}$.}
  \label{fig:argand}
\end{figure}

In order to disentangle the vertical structure we employ XPS-based techniques. XPS can be used to determine the relative coverages, if the atoms in each layer are chemically different. But, as expected, chemical shifts in the graphene components of the C~$1s$ core-level are found neither in T- nor in S-graphene (Fig.~\ref{fig:argand}(a)). We therefore apply the normal incidence x-ray standing wave (NIXSW) technique that combines dynamical x-ray diffraction with XPS and provides insights into both the vertical structure at the surface and layer coverages \cite{Zegenhagen2013, Zegenhagen1993, Woodruff1998, Woodruff2005}. In NIXSW each contribution to the XPS signal comes with a phase that is associated with the height of the corresponding photoemitter above the Bragg planes of the substrate. Contributions to the same XPS signal that originate from different heights can be therefore distinguished \cite{SuppMat}. 

In Fig.~\ref{fig:argand}(a), C~$1s$ core-level spectra are shown for the lowest (blue) and highest (red) preparation temperature for T- and S-graphene. The component at low binding energy can clearly be attributed to C in the SiC bulk, and the component at higher binding energy to C in \Gnul, \Gdrei or \ZL.  Hence, from the latter, only one C~$1s$ yield curve can be extracted (Fig.~\ref{fig:argand}(b)). Furthermore, since NIXSW identifies graphene layers by their different adsorption height (0th, 1st and 2nd layer) rather than by their azimuthal orientations, in this sum we have to designate the layers as ZL, \Gst and \Gnd, irrespective of their azimuthal orientation.  In the Argand diagram, the vector $\mathcal{D}^p_\mathrm{G}=F^p_\mathrm{c,G} e^{2 \pi i P^p_\mathrm{c,G}}$ of the measurement point $p$, with coherent fraction $F^p_\mathrm{c,G}$ and position $P^p_\mathrm{c,G}$, can be decomposed as the sum of the vectors corresponding to each layer weighted by the relative coverage pair $(\theta_\mathrm{G-1st}^p/\theta_\mathrm{ZL}^p;\theta_\mathrm{G-2nd}^p/\theta_\mathrm{ZL}^p)$ \cite{SuppMat}, i.e.,
\begin{equation}
	\mathcal{D}^p_\mathrm{G}=\frac{\mathcal{D}_\mathrm{ZL} + (\theta_\mathrm{G-1st}^p/\theta_\mathrm{ZL}^p)\mathcal{D}_\mathrm{G-1st}  + (\theta_\mathrm{G-2nd}^p/\theta_\mathrm{ZL}^p) \mathcal{D}_\mathrm{G-2nd}}{1+(\theta_\mathrm{G-1st}^p/\theta_\mathrm{ZL}^p)+(\theta_\mathrm{G-2nd}^p/\theta_\mathrm{ZL}^p)}
	\label{eq:three_layer_FT5}
\end{equation}
where we assume $\mathcal{D}_\mathrm{ZL}$ and $\mathcal{D}_\mathrm{G-1st}$ to be similar to EMLG and therefore take the experimentally measured values from Ref.~\cite{Sforzini2016}. For $\mathcal{D}_\mathrm{G-2nd}$, we further assume the distance \Gst and \Gnd  to be 3.37~\mbox{\AA} \cite{Nemec2013}, and the same coherent fraction than for $\mathcal{D}_\mathrm{G-1st}$, namely 1.  In the inset of Fig.~\ref{fig:argand}(c), we indicate by large black full circles the three vectors used in the sum, $\mathcal{D}_\mathrm{ZL}$, $\mathcal{D}_\mathrm{G-1st}$ and $\mathcal{D}_\mathrm{G-2nd}$ that are necessary to form the additional axes $\theta_\mathrm{G-1st}/\theta_\mathrm{ZL}$ and $\theta_\mathrm{G-2nd}/\theta_\mathrm{ZL}$ in Fig.~\ref{fig:argand}(c). The fit result $\mathcal{D}^p_\mathrm{G}$ of the yield curve is displayed in the Argand diagram in terms of ($F_\mathrm{c,G}^p; P_\mathrm{c,G}^p)$ and can be directly reread in term of the additional axes $(\theta_\mathrm{G-1st}^p/\theta_\mathrm{ZL}^p;\theta_\mathrm{G-2nd}^p/\theta_\mathrm{ZL}^p)$.

The colored triangles ($\medblacktriangledown$) are the sum vectors $\mathcal{D}^\mathrm{T-G}_\mathrm{G}$ measured on T-graphene obtained with different preparation temperatures (blue/red represents low/high preparation temperature), and the colored diamonds ($\medblackdiamond$) represent $\mathcal{D}^\mathrm{S-G}_\mathrm{G}$ for S-graphene. In agreement with ARPES results, we find that T-graphene is very inhomogeneous with $\theta^\mathrm{T-G}_\mathrm{G-1st}/\theta^\mathrm{T-G}_\mathrm{ZL}=0.6\pm0.1$ and $\theta^\mathrm{T-G}_\mathrm{G-2nd}/\theta^\mathrm{T-G}_\mathrm{ZL}$ varying from 0.1 to 0.6 depending on the preparation temperature. It shows that about 40\% of the \ZL remains always uncovered. With increasing preparation temperature, $\theta^\mathrm{T-G}_\mathrm{G-2nd}/\theta^\mathrm{T-G}_\mathrm{ZL}$ increases with the concomitant appearance of a $\pi$-band at $\mathrm{\bar K}_{30^\circ}$ in ARPES (Fig.~\ref{fig:ARPES}b), suggesting that areas solely covered by \ZL, single-layer \Gnul/\ZL and \tB/\ZL coexist on the surface. In contrast, S-graphene is much more homogeneous with $\theta^\mathrm{S-G}_\mathrm{G-1st}/\theta^\mathrm{S-G}_\mathrm{ZL} = 0.9 \pm 0.1$ and $\theta^\mathrm{S-G}_\mathrm{G-2nd}/\theta^\mathrm{S-G}_\mathrm{ZL} = 0.1 \pm 0.1$, independent of the preparation temperature. This means that only \Gnul is present on top of \ZL and that both have essentially the same coverage, while \Gdrei is absent. In conclusion, S-graphene is composed of a highly homogeneous \Gnul/\ZL, almost fully covering SiC(0001). Therefore, it constitutes an ideal platform to study \tB with different doping levels as well as different interaction strengths to the substrate by choosing the atomic species used for intercalation.

Finally, we briefly comment on the growth mechanism of graphene on SiC in a borazine atmosphere leading to the unconventional orientation $R0^\circ$. Because of the high temperature, \hBNnul does not stabilize \textit{in spite of} the presence of borazine. Yet, \textit{due to} the presence of the surfactant borazine molecule, the graphene layer that grows at this temperature is forced to adopt a lattice orientation of $R0^\circ$. The growth of this layer is a self-limiting process because the \ZL layer underneath \textit{is not} any more exposed to borazine and therefore grows in the orientation defined by the SiC substrate to which it \textit{is} exposed. We believe that this new preparation method, in which a surfactant enables the epitaxial growth of a graphene layer in an unconventional orientation, will foster new approaches to produce large-scale tBLG, thereby bringing its intriguing properties closer to applications.

\begin{acknowledgments}
  F.C.B., C.K. and F.S.T. acknowledge funding by the DFG through the SFB 1083 Structure and Dynamics of Internal Interfaces (project A 12). F.C.B thanks B. Amorim, M. Meissner, T. Seyller, S. Wolff, and M. \v{S}vec for fruitful discussion. We thank Diamond Light Source for access to beamline I09 (Proposals SI17737, SI20810, and SI20855) that contributed to the results presented here, and the I09 beam-line staff (P. K. Thakur, D. Duncan, and D. McCue) for their support during the experiment.
\end{acknowledgments}


\end{document}